# Online Peer-Assessment Datasets

**Michael Mogessie Ashenafi**

Department of Information Engineering and Computer Science, University of Trento, Italy

michael.mogessie@unitn.it

**Abstract:** Peer-assessment experiments were conducted among first and second year students at the University of Trento. The experiments spanned an entire semester and were conducted in five computer science courses between 2013 and 2016. Peer-assessment tasks included question and answer submission as well as answer evaluation tasks. The peer-assessment datasets are complimented by the final scores of participating students for each course. Teachers were involved in filtering out questions submitted by students on a weekly basis. Selected questions were then used in subsequent peer-assessment tasks. However, expert ratings are not included in the dataset. A major reason for this decision was that peer-assessment tasks were designed with minimal teacher supervision in mind. Arguments in favour of this approach are presented. The datasets are designed in a manner that would allow their utilization in a variety of experiments. They are reported as parsable data structures that, with intermediate processing, can be moulded into NLP or ML-ready datasets. Potential applications of interest include performance prediction and text similarity tasks.

**Keywords:** peer-assessment, dataset, performance prediction, automated answer scoring

## 1. INTRODUCTION

Peer-assessment in education is an assessment method in which students assess the performance of their peers. Topping (1998) defines peer-assessment more formally as Topping (1998) defines peer-assessment more formally as "… an arrangement in which individuals consider the amount, level, value, worth, quality, or success of the products or outcomes of learning of peers of similar status."

A wide variety of peer-assessment settings exist in which the nature of the work being assessed varies with the discipline and course. Essays, answers to open questions, and oral presentations are examples of work that is assessed in peer-assessment classes.

Reliability, validity, and practicality of peer-assessment as well as its impact on students' learning have been studied for decades. Nonetheless, there is no strong consensus among practitioners on whether peer-assessment is guaranteed to deliver its desired effects.

Validity and reliability of peer-assessment remain the two main issues with the practice, which have perhaps limited its adoption as an alternative method of assessment. Indeed, most peer-assessment studies were conducted to determine if student-assigned grades were good enough to be considered equivalent to teacher grades (Ashenafi, 2017; Falchikov & Goldfinch, 1998). Despite continued research in this area, the consensus is that whoever chooses to adopt peer-assessment, they should proceed with caution and carefully consider the many variables that may that may affect the reliability and validity of peer grades.

The wide range of scenarios in which peer-assessment is implemented has also made it difficult to reach solid conclusions about its effectiveness as the number and impact of variables being studied varies from one scenario to another. Some variables are however common to all peer-assessment settings. Examples include the number of students involved per assessment task and the total number of participants.

Online forms of peer-assessment are now more common than before, at least among instructors seeking to introduce the practice in their classrooms in an efficient manner. This study presents peer-assessment datasets obtained from an online platform used in several undergraduate-level computer science courses. A description of manual and automated or online peer-assessment is presented and the advantages of automating peer-assessment practices are outlined. A case for open research and public peer-assessment datasets is also made. Then, a thorough description of the datasets is provided.

## 2. MANUAL VERSUS AUTOMATED OR ONLINE PEER-ASSESSMENT

A review of recent research in peer-assessment would tell us that topics of interest differ for researchers in education and pedagogy and their counterparts in computer science and engineering fields. For those interested in pedagogy and educational psychology, focus is more on themes such as peer-feedback, design strategies, student perceptions, social and psychological factors and validity and reliability of the practice (Ashenafi, 2017).

For researchers in computer science and engineering, research is driven by data collected from online educational platforms such as Massive Open Online Courses (MOOC) and other peer-assessment frameworks. Here focus is more on how to improve efficiency of conducting peer-assessment tasks, how to account for bias and other errors in judgment and how to calibrate peer grades (Pare & Joordens, 2008; Goldin, 2012; Piech et al., 2013; Suen, 2014).

Regardless of the topic of interest, researchers commit to one of two forms of peer-assessment as the decide the environment in which to conduct experiments. They either choose to use peer-assessment in a traditional classroom environment, without the use of educational technologies, or adopt a computer-based approach, which usually involves using an online peer-assessment framework.

These distinctions between these choices calls for drawing the line between manual, traditional forms of peer-assessment and peer-assessment that is conducted in a technology-supported setting. Formal definitions of the two cases are provided below.

**Traditional (Manual) Peer-Assessment** practices are those that do not utilize electronic equipment such as Electronic Voting Systems (EVS) or clickers or information technology artefacts such as computer software to improve the efficiency and effectiveness of processes. In such practices, the work to be assessed is either written or orally presented. The collection and assignment of the work to be assessed entails the teacher manually collecting students' works and redistributing them, whether randomly or in a pre-determined manner. Students rate and comment on their peers' works either by providing feedback written on paper or orally. The specification and communication of criteria, if any, takes place in the form of traditional classroom discussions or lessons.

**Automated (Online) Peer-Assessment**, on the other hand, utilizes electronic equipment or information technology artefacts to automate, partly or entirely, the processes involved. Typical automated peer-assessment uses computer software or Internet technologies or both to facilitate the distribution and delivery of assessment tasks as well as the submission of tasks and communication of assessment results. Some automated peer-assessment tools also provide teachers the choice to specify assessment criteria to be used by students when assessing their peers' works. Semi-automated peer-assessment refers to practices in which only certain tasks are automated.

Conducting peer-assessment in a traditional or manual manner introduces factors that have the potential to impact the effectiveness and efficiency of the practice. Indeed, most research in peer-assessment focuses on determining whether these challenges render the practice inefficient or even ineffective (Falchikov & Goldfinch, 2000; Ashenafi, 2017). One of the cases for automated or online peer-assessment arises from its potential to reduce or eliminate these effects with a reasonable amount of effort. The most common challenges of manual peer-assessment and how automated peer-assessment may tackle them are discussed below.

## 2.1 Disclosure of Identity

Naturally, peer-assessment tasks are conducted in open classroom environments. Students are familiar with each other and it is not difficult for a student to learn whose work they are reviewing and who is reviewing their work. The teacher can introduce anonymity by having each student write unique identifiers instead of their names on the work to be assessed. But this implies the extra work of assigning identifiers to students and mapping students' names to their respective identifiers. The workload on the instructor grows significantly with increase in the number of students as well as the number of assessors per assessment task. For this reason, some teachers choose not to introduce anonymity into the peer-assessment process. This in turn leads to issues that may have negative impact on peer-assessment. Bias and favoritism are the most common examples of such issues.

Unintended disclosure of identity information is one of the problems that are easily solved by automating the peer-assessment process. Implementing anonymity in automated peer-assessment software does not require any special algorithm and is as easy as withholding any identity information from the assessor. Increase in the number of students has virtually no effect on how the software system implements anonymity either.

## 2.2 Increase in Teacher Workload

In a manual peer-assessment environment, the teacher must collect assignments, redistribute them, and collect assessment results in a manual manner. The teacher's tasks become more complex if assignments are to be distributed randomly and if more than one student is to assess another student's work. Eventually, most of the teacher's time will be spent on technical tasks rather than actual teaching activities. The workload quickly becomes restrictive with small increase in the number of students. Therefore, manual peer-assessment is not scalable and the impacts discussed here can only be mitigated when it is conducted in a class with few students.

Automated peer-assessment can greatly reduce the teacher's workload because almost all laborious tasks can be automated. Random distribution of tasks does not require sophisticated algorithms. Algorithms can be developed that handle additional constraints and requirements from the teacher on how to assign tasks. For instance, the teacher may decide that assignments should be randomly distributed while making sure students are not assigned the works of peers they had assessed previously. A hypothetical situation in which a teacher intends to randomly distribute assignments from twenty students with the condition that each assignment must be assessed by three peers and that each peer must assess only two assignments should be sufficient to appreciate how daunting manual distribution of assignments can be. Efficient algorithms that require a few seconds or less to complete such tasks can be developed in accordance with specific requirements.

## 2.3 Academic Dishonesty

In a manual peer-assessment environment, scalability issues prevent efficient detection of academic dishonesty such as plagiarism. Detection of other dishonest behavior such as formation of cliques to award marks reciprocally becomes impossible as the number of students involved increases.

Although transition to the automated peer-assessment realm does not eliminate academic dishonesty, it greatly improves detection of such behavior by providing the platform to apply state-of-the-art solutions in computer science that deal with such issues.
Plagiarism detection capability of software tools has improved with advances in disciplines within computer science such as Information Retrieval (IR), Natural Language Processing (NLP), and Machine Learning (ML).

Plagiarism detection software uses one or a combination of these techniques to measure the similarity of work submitted by students or authors. In peer-assessment scenarios, such detection could either involve comparing the similarity of submissions within a class or comparing students' works with literature from external sources.

A variety of plagiarism detection tools, both free and commercial, exist. Turnitin[1] is a commercial plagiarism detection tool in use by many higher education institutions. Crosscheck[2] is another plagiarism detection service used by many academic journals and publishers.

---

[1] http://turnitin.com/
[2] http://www.ithenticate.com/products/crosscheck

By integrating plagiarism detection into automated peer-assessment, practitioners can delegate computer software to monitor such dishonest behavior.

Systematic cheating behavior such as reciprocity and the formation of cliques in peer-assessment scenarios can also be detected using advanced computer algorithms. A class implementing peer-assessment may be regarded as a social network that emerges from the interactions of students during assessment processes. Peer-assessment tasks performed in multiple cycles of assignment distribution and submission give rise to more complex social networks. In classes implementing manual peer-assessment, it is impossible to model and analyze such networks. In automated peer-assessment scenarios, models of social networks that emerge from student interactions may be constructed and analyzed using advanced algorithms. Graph theory and social network analysis are well studied problems. Digital social networks such as Facebook, Google+, and Twitter already use such algorithms to identify important relationships among users. Therefore, a transition to automated peer-assessment allows taking advantage of advances in social network analysis to investigate intricate student relationships such as the existence of cliques of students who may engage in academic dishonesty practices such as collusion.

## 2.4 Manual Grading of Assignments

To measure the degree of agreement between teacher and peer-assigned scores, teachers must score each assessment task that has been scored by students. The time needed to manually grade assignments increases with increase in the number of assessment tasks and hence, with increase in the number of students involved in the peer-assessment experiment. The number of peer-assessment cycles and the content of assignments determine teacher workload as well.

An analysis of 12 teacher-peer score agreement studies conducted since 2000 shows the number of assignments corrected by teachers ranged from 15 to 272 (Ashenafi 2017). Obtaining representative sample sizes in large classes therefore requires manual correction of a significantly large number of assignments.

Automated essay scoring is an alternative to manual scoring, which uses Natural Language Processing (NLP), Machine Learning (ML) techniques or a combination of both to analyze and assign scores to students' essays. Automated essay scoring has improved over the years with advances in NLP and ML. For instance, Educational Testing Service (ETS) has developed an automatic essay scoring computer program, which it uses to verify the accuracy of human readers that score the essays of test takers (Attali & Burstein, 2006).

Automating peer-assessment tasks may therefore greatly reduce the number of teacher-graded essays through automated essay scoring. A direct consequence of applying automated essay scoring in peer-assessment environments is the ability to involve as many students as needed without imposing additional workload on the teacher, which implies that larger sample sizes can be utilized with minimum effort on the part of practitioners to improve the quality of teacher-peer score agreement studies.

The question of validity and reliability of peer-assessment has been a deterring factor in the adoption of the practice. An interesting question regarding teacher-peer score agreements is whether educators would willingly adopt peer-assessment, either as a formative or summative assessment tool, if research proved that in properly designed peer-assessment practices peer-assigned scores have strong correlation, or even equivalence, with teacher-assigned scores.

One way to trigger positive responses to this important question is to engage in large-scale research into validity and reliability of peer-assigned scores. Automation of peer-assessment tasks empowers researchers with software tools in their quest to determine whether student assessors can be used as substitutes for teachers, not only by providing the platform to conduct large-scale teacher-peer score agreement experiments but also through automated calibration of scores to resemble those of teachers (Pare & Joordens, 2008; Goldin, 2012; Piech et al., 2013).

## 3. SOME ONLINE PEER-ASSESSMENT FRAMEWORKS

There is an extensive number of tools that support automated peer-assessment. However, only tools that the authors consider to be representatives of how many other frameworks commonly model peer-assessment tasks are presented here. See Luxton-Reilly (2009) for a comprehensive review of tools that support peer-assessment.

### 3.1 PRAISE (Peer Review Assignments Increase Student Experience)

de Raadt et al. (2009) presented a generic peer-assessment tool that was used in the fields of computer science, accounting and nursing. The instructor could specify criteria before distributing assignments, which students would use to rate their peers' assignments. The system could compare reviews and suggest a mark. Disagreements among reviewers would lead the system to submit the solution to the instructor for moderation. The instructor would then decide the final mark. The system facilitated anonymity and promoted feedback from multiple sources.

### 3.2 PeerWise

Denny et al. (2008) presented PeerWise, a peer-assessment tool, which students used to create multiple-choice questions and answer those created by their peers. When answering a question, students would also be required to rate the quality of the question. They could also comment on the question, in which case the author of the question could reply to the comment.

### 3.3 PeerScholar

Paré and Joordens (2008) presented another peer-assessment tool, which was initially designed with the aim of improving writing and critical thinking skills of psychology students. First, students would submit essays. Next, they would be asked to anonymously assess the works of their peers, assign marks between 1 and 10, and comment on their assessments. An additional feature of PeerScholar is that students could also rate the reviews the received.

### 3.4 Calibrated Peer Review (CPR)

The Calibrated Peer Review (CPR) system (Russell, 2004) is a web-based framework that facilitates submission and review of written assignments. CPR is course independent but requires students be trained before engaging in peer reviews. CPR decouples the review process from class size, making it applicable in courses with a large number of students. In CPR, students evaluate their own work as well. The system delivers significant amount of feedback to the instructor about student performance throughout the review process.

### 3.5 Aropä

Aropä (Hamer et al., 2007) is a web-based peer assessment support tool which supports manual or automatic allocation of reviews. Students review their peers' work both qualitatively and quantitatively. Quantitative review follows rubrics and criteria. As in PeerScholar, reviewers themselves may be rated. Instructors provide review weights before final marks based on average review marks can be assigned.

Babik et al. (2016) presented a taxonomy framework that can be used to categorize online peer-assessment systems. The taxonomy allows analyzing online peer-assessment frameworks based on how they elicit evaluation, how peer-assessment results are reported to students and the instructor, how peer-assessment processes are structured, how issues such as inaccuracy and bias are addressed and if and how the peer-assessment systems promote higher-level learning and other benefits. Given that there are currently many peer-assessment tools developed in-house, such taxonomy would help researchers intending to move their experiments to online platforms.

Recent online peer-assessment studies focused on exploring psychological and cognitive aspects of the practice such as conceptions and approaches to learning (Yang & Tsai, 2010), feedback and improved learning (Chen & Tsai, 2009) and its formative values (Mostert & Snowball, 2010).

Some studies have gone further and utilized data collected from online peer-assessment platforms to build mathematical models that could calibrate peer grades and account for factors such as student bias and other errors in judgment (Pare & Joordens, 2008; Goldin, 2012; Piech et al., 2013). Such research results indicate that use of technology improves the effectiveness of the practice and that applying machine learning to current issues in the practice is certainly a promising direction.

### 4. THERE IS MORE TO ONLINE PEER-ASSESSMENT

In many of the online peer-assessment systems discussed here, the advantages of online peer-assessment hardly go beyond improving efficiency. Indeed, take away efficiency and what would remain is only an electronic version of manual peer-assessment processes. The main argument here is that online peer-assessment has more to offer than just efficiency, which may also come in the form of improved data collection.

In addition to addressing the challenges of manual peer-assessment discussed earlier, automation provides new opportunities for both students and teachers. Three immediate opportunities arising from automation of peer-assessment tasks are discussed here.

## 4.1 Student modeling

When involved in multiple cycles of peer-assessment, students generate significant amounts data, which can be used to build representative models. Such student models are important in investigating student progress. One way in which a student model can be used is to identify groups of students with similar levels of performance. The teacher could then use this information to facilitate partitioning the class when assigning group projects.

Machine learning algorithms can also be applied to students' previous exam results and student models constructed using data from the respective course to predict performance of students enrolled in new editions of the course. Performance prediction, when performed at various stages of a course, will serve as an early intervention mechanism. In this manner, automated peer-assessment can be used to monitor student progress and to identify students that may require special supervision (Ashenafi et al., 2016).

## 4.2 Mobility

Some peer-assessment tasks can be performed either in-class or out-of-class. Some may be designated as take-home assignments. Automation of such tasks provides the opportunity to introduce mobility into the peer-assessment experience. Peer-assessment software solutions can be developed for smartphones and tablets, making the experience appealing to today's student society.

Some peer-assessment tasks need not be completed in-class. Students will then have the option to complete such tasks while on-the-move. This opens the door to the automated collection of even more data about students, contributing to the construction of robust student models.

## 4.3 Setting the Stage for Further Research

Automation enables researchers to conduct studies that could otherwise be difficult or impossible to conduct in manual peer-assessment environments. An example is an experiment that attempts to establish whether students' criticism of their peers' abilities to play the role of assessors originates from bias.

In automated peer-assessment scenarios, where anonymity is maintained, the teacher or a group of teachers can play the role of a student and perform assessment tasks, without being identified by "peers". After enough number of assessments, student opinions can be obtained via questionnaires or interviews to determine if such criticisms have any foundation.

Automation also allows efficient adoption of practices from well-advanced disciplines. For instance, applying game theory to peer-assessment tasks could increase student engagement by introducing healthy competition into the practice.

It should be considered that peer-assessment tasks have intrinsic game-like features, which, in theory, students might be able to exploit by adopting specific strategies. An example is where a student adopts the strategy of assigning low marks and providing poor feedback to other students to improve their ranking. The complexity of game theory is possibly reflected in this scenario if all students adopted this same strategy, which would lead to all students earning unfairly low marks and beats the purpose of the practice.

Although this is a theoretical scenario, similar behavior in intelligent tutoring systems, referred to as "gaming the system", has been identified (Baker et al., 2004). Advanced computer algorithms may be developed to support detection of such unintended behavior in classes implementing automated peer-assessment.

There is a significant requirement, however, that must be met before many of the benefits of online peer-assessment outlined here can be ripped and its potential reached. The research community must move to an open peer-assessment practice that would allow conducting open and reproducible experiments. This essentially implies making peer-assessment data openly available for fellow and prospective researchers.

Unfortunately, the availability of open peer-assessment datasets is modest, at best. Currently, the authors could obtain only one peer-assessment dataset published by Vozniuk et al. (2016). The dataset contains report grades assigned by students as well as teachers but does not contain the reports themselves. This dataset is hence suitable for peer-assessment validity and reliability experiments only.

The authors believe that peer-assessment data may contribute not only to validity and reliability studies but also to studies that explore topics such as student performance prediction and automated essay scoring.

The discussions so far were intended to set the stage for the presentation of the peer-assessment datasets and to highlight how they may promote open and incremental peer-assessment research.

The way a peer-assessment system is designed determines what type of data are collected and how the dataset may be constructed. A discussion of how peer-assessment tasks were designed and what type of data were collected thus follows.

## 5. THE ONLINE PEER-ASSESSMENT SYSTEM

A web-based peer-assessment platform was developed at the University of Trento and used in five undergraduate-level computer science courses between 2012 and 2016. All courses were conducted in

Italian. Participation in peer-assessment tasks was optional. However, all students who completed at least a third of the tasks were awarded a bonus worth 3.3% of the final mark. An additional 3.3% bonus was awarded to the top-third students, based on the number of peer-awarded points. For all five courses, it was observed that active participation in peer-assessment activities dropped towards the end of the course.

The peer-assessment platform was revised after its use in the first two courses. Students completed a weekly peer-assessment cycle composed of three tasks. First, students submitted questions relating to a list of previously discussed topics provided by the teacher. In an intermediate task, students rated the reliability, interestingness and difficulty levels of the questions. Then, the teacher examined the questions and selected a subset of questions, which were randomly assigned to students in the second task. The assignment procedure ensured that each question was assigned to at least four students. Machine learning techniques were applied to group similar questions in an effort to facilitate the question selection process. Students then proceed to submit their answers.

After answers were submitted, question-answer sets were randomly distributed to students. In the first version of the peer-assessment system, students voted for the answer they perceived the best. In the second version, however, students rated each answer. To encourage careful assessment of each answer, students were provided with a certain amount of points, referred to as coins, to distribute over the answers. The number of coins available for distribution was computed from the number of answers submitted to that question.

After the completion of each cycle, questions and answers, together with coins earned, were made available to all students. Students could also monitor their progress by accessing visual and statistical information available in their profile page.

## 6. THE DATASETS

Separate datasets were constructed for the five courses. The version of the peer-assessment system is reported as the course version. Version 1 courses use simple votes whereas version 2 courses use coin-based rating of answers. Students' course grades are also included in the datasets. The Italian higher education system uses a 0-30 grading scale, with 18 as the minimum passing score. For those students who did not complete the course successfully but participated in peer-assessment tasks regardless, a score of 0 is reported.

It is worth stressing that peer-assessment tasks were designed with minimal teacher supervision in mind. This is evident in that none of the answers have been assigned teacher grades. While the datasets may not directly be used in peer-assessment validity experiments, they can certainly be used to build models that explore whether such validity may be inferred from course grades assigned by the teacher.

The decision to make participation in peer-assessment tasks non-compulsory is reflected by the fact that some answers were assessed by a fewer number of peers than others. It is also worth noting that task incompletion rates increased towards the final weeks for all five courses. Despite this, a total of 83% of students for three of the courses completed at least a third of the tasks. Details of the peer-assessment system and arguments for the approach followed in designing peer-assessment tasks are reported in Ashenafi et al. (2014).

## 6.1 Dataset Structure

Because weekly peer-assessment tasks started with the submission of questions, a subset of which were used as inputs to subsequent tasks, it was decided to structure the datasets in a similar manner. Every course consisted of lectures, which in turn were composed of questions submitted in the first task of the week, "Ask A Question". Question attributes such as the number of evaluations and ratings of difficulty, relevance and interestingness are reported. For questions that were not evaluated, 0 values are reported.

The question text, information about the student who submitted it and, whenever available, answers are also reported. Each answer structure contains the answer text, the student who provided it and its peer-ratings. Depending on the version of the course, which is also reported as a course attribute, this rating may be reported as a simple vote or as a set of coins awarded to the answer. For every student that provided an answer to a question, their course grade is reported as well.

## 6.2 Metadata

The complete structure of the datasets is presented in table 1 and an explanation for each attribute of the dataset is provided in table 2. One dataset per course is provided. Over 4800 questions and over 5000 answers were submitted by over 800 students that enrolled in the five courses between 2013 and 2016. A breakdown is provided in table 3.

The datasets are formatted as JavaScript Open Notation (JSON) objects. A variety of programming languages support, natively or via the use of external libraries, parsing JSON objects. A Java library that can readily parse the datasets into Java objects is provided together with the datasets. It is hoped that, with intermediate processing, the JSON files can be streamlined and transformed into datasets that can be used in several machine learning and NLP tasks.

**Table 1: Structure of the datasets**

| Course | Lecture | Question |
|---|---|---|
| {<br>"courseId":Integer,<br>"version":Integer,<br>"courseName": String,<br>"lectures":[**Lecture**]<br>} | {<br>"lectureId":Integer,<br>"lectureTitle:String,<br>"questions":[**Question**]<br>} | {<br>"questionId":Integer,<br>"asker":**Asker**,<br>"task":**Task**,<br>"questionText":String,<br>"totalDifficultyLevel":Integer,<br>"totalInterestingnessLevel":Integer,<br>"totalRelevanceLevel":Integer,<br>"numEvaluators":Integer,<br>"chosenForAnswering":boolean, |

|  |  | "chosenForMultipleChoice":boolean, "keywords":[**Keyword**], "notes":String, "answers":[**Answer**] } |
|---|---|---|
| **Asker** | **Task** | **Keyword** |
| { "courseId":Integer, "askerId":Integer, "courseFinalScore":Integer } | { "taskId":Integer, "taskName":Integer } | { "keyword":String } |
| **Answer** | **Responder** | **Coin** |
| { "answerId":Integer, "task":**Task**, "responder":**Responder**, "answerText":String, "notes":String, "rating": Integer, "coins": [**Coin**] } | { "courseId":Integer, "reponderId":Integer, "courseFinalScore":Integer } | { "coinId":Integer, "rater":**Rater**, "task":**Task**, "value":Integer } |
| **Rater** |  |  |
| { "courseId":Integer, "raterId":Integer, "courseFinalScore":Integer } |  |  |

**Table 2: Description of Dataset Attributes with Primitive Datatypes – Integer, String and Boolean**

| Name | Type | Description |
|---|---|---|
| courseId | Integer | The course's unique identifier |
| version | Integer | The version of the system used, either 1 or 2 |
| courseName | String | Name of the course |
| lectureId | Integer | The lecture's unique identifier |
| lectureTitle | String | The lecture's title |
| questionId | Integer | The question's unique identifier |
| askerId | Integer | Id of the student who asked the question |
| courseFinalScore | Integer | The student's final score for the course |
| taskId | Integer | The task's unique Identifier |
| taskName | String | The task's name, E.g. "Ask A Question" |
| questionText | String | The text of the question |
| totalDifficultyLevel | Integer | The question's difficulty as rated by students |

| totalRelevanceLevel | Integer | The questions relevance as rated by students |
| totalInterestingnessLevel | Integer | The questions interestingness as rated by students |
| numEvaluators | Integer | Number of students who evaluated the question |
| chosenForAnswering | Boolean | True if question was used in 'Answer A Question' tasks |
| chosenForMultipleChoice | Boolean | True if question was used in multiple choice or rating tasks |
| Keyword | String | A keyword provided when submitting the question |
| Notes | String | Optional notes provided when submitting a task |
| answerId | Integer | The answer's unique identifier |
| responderId | Integer | Id of the student who provided the answer |
| answerText | String | The text of the answer |
| rating | Integer | The answer's rating, used only in version 1 courses |
| coinId | Integer | A unique identifier of a coin or point awarded to an answer |
| raterId | Integer | A unique identifier of the student who awarded the coin |
| Value | Integer | The coin's value |

**Table 3: Additional Course Metadata**

| Course Name | Questions | Answers | Students | Dataset Filename |
|---|---|---|---|---|
| Informatica Generale 1 | 1303 | 1398 | 204 | 2_ig1.json |
| Programmazione 2 | 1013 | 1041 | 163 | 4_pr2.json |
| Programmazione 1 | 547 | 728 | 132 | 5_pr1.json |
| Linguaggi Programmazione 1 | 1087 | 1146 | 179 | 100_lp1.json |
| Lingauggi Programmazione 2 | 858 | 972 | 183 | 102_lp2.json |

## 6.3 How to Obtain the Datasets

The dataset files and the java parser library are freely available at https://github.com/michaelmogessie/t4e_datasets. All personally identifiable information has been removed from the datasets.

## 7. DISCUSSION AND CONCLUSION

Challenges preventing large-scale and extensive studies in peer-assessment that arise from the manual nature of tasks were discussed. It was argued how automation of peer-assessment tasks may help researchers focus on problems that have more to do with than just efficiency. A semi-automated peer-assessment prototype was presented to provide insight into how the datasets reported in this study were constructed.

Peer-assessment research can greatly benefit from automation of tasks as it will be possible to apply machine learning and natural language processing techniques to tackle problems that are not inherent in the discipline itself. There is also promise in automated peer-assessment to extend the practice to student supervision and to use it as an early intervention mechanism.

Large amounts of peer-assessment data may give back to research in the practice itself, such as large scale validity and reliability studies as well as bring learning analytics to peer-assessment. Student performance prediction, automated essay scoring and domain specific Question Answering studies may all benefit from peer-assessment data.

The purpose of the datasets presented here is to promote such studies. None of the frameworks or studies discussed earlier make their peer-assessment data openly available. This may present a challenge for those that aim to conduct research based on the findings of those studies.

An example is research that aims to investigate the correlation between student perception and performance in peer-assessment tasks or final exams. Although most studies report results such as overall student perception obtained from survey results and selected responses or feedbacks to peer-assessment tasks, they do not publish the data in their entirety. Hence, conducting experiments that investigate the existence of correlations between any of these factors would require using an online peer-assessment system to gather similar data.

The authors believe that making these datasets publicly available paves the way to peer-assessment experiments that are replicable and extensible.

The datasets contain not only information about peer-assessment experiments but also question and answer texts that may be used in Italian NLP tasks such as Question Answering and Automated Essay Scoring. Previous experiments using some of these datasets have demonstrated the promising potential of peer-assessment in predicting student success and modelling progress (Ashenafi et al., 2015; 2016). The datasets were therefore constructed with not a specific experiment in mind. However, it is hoped that their representation allows extraction of only required pieces of information with little effort.

Whether earning higher peer-marks for answering questions with higher levels of difficulty is correlated with higher final scores or if participation in peer-assessment tasks leads to successful course completion are examples of research questions that may be explored using these datasets.

Task incompletion was one of the challenges that were faced in all rounds of peer-assessment in the framework used in this study. It has influenced the completeness of the datasets and, to a degree, the fairness of points earned by students who participated in online peer-assessment tasks. Naturally, students who are assigned answer-rating tasks but choose not to complete them take away points from students who have responded to questions. Research on remedies for this issue may also be conducted using these datasets.

Peer-assessment has been utilized in all levels of education but has yet to take advantage of advances in computer science an information and communication technologies. It is hoped that the availability of these datasets will promote further research in issues in peer-assessment and demonstrate its potential as a technology-supported educational discipline.